\documentstyle[12pt]{article}
\begin{document}
\baselineskip=16pt
\newcommand{\newc}{\newcommand}
\newc{\be}{\begin{equation}}
\newc{\ee}{\end{equation}}
\newc{\bea}{\begin{eqnarray}}
\newc{\eea}{\end{eqnarray}}
\newc{\bean}{\begin{eqnarray*}}
\newc{\eean}{\end{eqnarray*}}
\newc{\eqn}[1]{(\ref{#1})}
\newc{\gsim}{\lower.7ex\hbox{$\;\stackrel{\textstyle>}{\sim}\;$}}
\newc{\lsim}{\lower.7ex\hbox{$\;\stackrel{\textstyle<}{\sim}\;$}}
\newc{\ra}{\rightarrow}
\newc{\tanb}{\tan\beta}

\newc{\mz}{m_Z}
\newc{\hl}{h}	\newc{\hh}{H}
\newc{\ha}{A}	\newc{\hcpm}{H^\pm}
\newc{\mhl}{m_h}	\newc{\mhh}{m_H}	\newc{\mha}{m_A}
	\newc{\mhcpm}{m_{\hcpm}}
\newc{\rhocrit}{\rho_{\rm crit}}
\newc{\abund}{\Omega h_0^2}

\newc{\neutone}{\chi^0_1}
\newc{\neuttwo}{\chi^0_2}
\newc{\mneutone}{m_{\neutone}}	\newc{\mneuttwo}{m_{\neuttwo}}
\newc{\mcharone}{m_{\charone}}	\newc{\charone}{\chi_1^\pm}

\newc{\ie}{{\it i.e.}}
\newc{\etal}{{\it et al.}}
\newc{\eg}{{\it e.g.}}
\newc{\etc}{{\it etc.}}
\newc{\sinsqthw}{\sin^2\theta_{\rm w}}
\newc{\costhw}{\cos\theta_{\rm w}}
\newc{\cossqthw}{\cos^2\theta_{\rm w}}
\newc{\costhwfourth}{\cos^4\theta_{\rm w}}

\newc{\ev}{{\rm\,eV}}
\newc{\mev}{{\rm\,MeV}}
\newc{\gev}{{\rm\,GeV}}
\newc{\tev}{{\rm\,TeV}}
\newc{\vev}{\hbox{\it v.e.v.}}
\newc{\vone}{v_d}	\newc{\vtwo}{v_u}
\newc{\higgsino}{{\tilde{H}}}

\def\thefiglist#1{\section*{Figure Captions\markboth
{FIGURE CAPTIONS} {FIGURE CAPTIONS}}\list
{Figure \arabic{enumi}.}
{\settowidth\labelwidth{Figure #1.}\leftmargin\labelwidth
\advance\leftmargin\labelsep
\usecounter{enumi}}
\def\newblock{\hskip .11em plus .33em minus -.07em}
\sloppy}
\let\endthefiglist=\endlist
\def\NPB#1#2#3{Nucl. Phys. {\bf B#1} (19#2) #3}
\def\PLB#1#2#3{Phys. Lett. {\bf B#1} (19#2) #3}
\def\PLBold#1#2#3{Phys. Lett. {\bf#1B} (19#2) #3}
\def\PRD#1#2#3{Phys. Rev. {\bf D#1} (19#2) #3}
\def\PRL#1#2#3{Phys. Rev. Lett. {\bf#1} (19#2) #3}
\def\PRT#1#2#3{Phys. Rep. {\bf#1} (19#2) #3}
\def\ARAA#1#2#3{Ann. Rev. Astron. Astrophys. {\bf#1} (19#2) #3}
\def\ARNP#1#2#3{Ann. Rev. Nucl. Part. Sci. {\bf#1} (19#2) #3}
\def\MODA#1#2#3{Mod. Phys. Lett. {\bf A#1} (19#2) #3}
\def\ZPC#1#2#3{Zeit. f\"ur Physik {\bf C#1} (19#2) #3}
\def\APJ#1#2#3{Ap. J. {\bf#1} (19#2) #3}

\newc{\mchi}{m_\chi}	\newc{\mx}{\mchi}
\newc{\mi}{m_{\chi^0_i}}	\newc{\mj}{m_{\chi^0_j}}
\newc{\pone}{p_1}	\newc{\ptwo}{p_2}
\newc{\kone}{k_1}	\newc{\ktwo}{k_2}
\newc{\vecpone}{\vec\pone}	\newc{\vecptwo}{\vec\ptwo}
\newc{\veckone}{\vec\kone}	\newc{\vecktwo}{\vec\ktwo}
\newc{\Eone}{E_1}	\newc{\Etwo}{E_2}
\newc{\mf}{m_f}
\newc{\mF}{m_F}
\newc{\cth}{\cos\theta}
\newc{\tminus}{t_-}	\newc{\tplus}{t_+}	\newc{\tpm}{t_{\pm}}
\newc{\tzero}{t_0}
\newc{\deltat}{\Delta t}
\newc{\vrel}{v_{\rm rel}}
\newc{\sigmav}{\sigma\vrel}
\newc{\azero}{a^0}	\newc{\bzero}{b^0}
\newc{\mbarsquare}{|\overline{\cal M}|^2}
\newc{\deli}{\Delta_{\chi_i^0}}	\newc{\delj}{\Delta_{\chi_j^0}}
\newc{\abundchi}{\Omega_\chi h_0^2}
\newc{\abundtot}{\Omega_{\rm TOT} h_0^2}
\newc{\bino}{\widetilde B^0}
\newc{\wino}{\widetilde W^0_3}
\newc{\hinob}{\widetilde H_b}       \newc{\hinot}{\widetilde H_t}
\newc{\hinos}{\widetilde H_S}       \newc{\hinoa}{\widetilde H_A}
\newc{\mtwo}{M_2}

\begin{titlepage}
\begin{flushright}
{\large
UM-TH-94-02\\
NSF-ITP-94-27\\
hep-ph/9404227\\
February 1994\\
}
\end{flushright}
\vskip 2cm
\begin{center}
{\large\bf
A Simple Way of Calculating Cosmological Relic Density
}	
\vskip 1cm
{\Large
Leszek Roszkowski\footnote{
E-mail address: {\tt leszek@leszek.physics.lsa.umich.edu}
}
\\}
\vskip 2pt
{\large\it Randall Physics Laboratory\\ University of Michigan\\ Ann
Arbor,
MI 48190, USA}\\
\end{center}
\vskip .5cm
\begin{abstract}
A simple procedure is presented which leads to a dramatic
simplification in the
calculation of the relic density of stable particles in the Universe.
\end{abstract}
\end{titlepage}
\setcounter{footnote}{0}
\setcounter{page}{2}
\setcounter{section}{0}
\setcounter{subsection}{0}
\setcounter{subsubsection}{0}
\newpage

Any stable species $\chi$ contributes to the total mass-energy
density in the
Universe. If its number density cannot be reduced
efficiently enough before it decouples from the thermal equilibrium,
its relic abundance $\abundchi$ can be sizeable and it can affect the
evolution and the age of the Universe. A conservative estimate that
the
Universe is at least 10 billion years old requires
$\abundchi<\abundtot<1$~\cite{kt}. Furthermore, there is growing
evidence for
dark matter at both galactic and larger length scales~\cite{kt} which
would
most likely require the existence of some
type of exotic neutral particles. Since such particles
are often present in many theories beyond the Standard Model,
it is important to develop a simple practical procedure for
calculating their
relic abundance with enough precision.

Two groups~\cite{swo,gg} have developed equivalent
frameworks for properly calculating the relic density of $\chi$'s,
including
relativistic corrections. The method of Ref.~\cite{swo} is in
practice
applicable away from poles and new final-state thresholds, which is
most often
the case.
In Ref.~\cite{gg} also the vicinity of poles and thresholds
has been carefully studied. Essentially, one needs to calculate
the thermally averaged product of the $\chi\bar\chi$ annihilation
cross section
and their relative velocity $\langle \sigmav \rangle$. In practice,
one expands
$\langle \sigmav \rangle=a+bx+ {\cal O}(x^2)$
in powers of $x\equiv T/\mchi={\cal O}(1/20)$ in order to avoid
difficult
numerical integrations and approximates $\langle \sigmav \rangle$ by
$a$ and
$b$.
Both techniques give equivalent results~\cite{gg,msprivate}
in the overlapping region (away from poles and thresholds).

Unfortunately, in practice the actual calculation of even the first
two terms
of
the expansion is typically very complicated and tedious.
In this Letter I report on a dramatic technical simplification in
practical
applications of the method of Ref.~\cite{swo}.

Consider an annihilation of particles $\chi$, $\bar\chi$ into a
two-body final
state. Furthermore, in many cases of interest, the final state
particles have
equal mass $\mF$.
(A general case of unequal final state masses will be presented
elsewhere~\cite{lrinprep}.)
Let the momenta of the two initial states $\chi$, $\bar\chi$, and the
two final
states $F$, $\bar F$, be $\pone$, $\ptwo$ and $\kone$, $\ktwo$,
respectively.
Srednicki, \etal, introduce the function $w(s)$ defined as
\be
w(s)\equiv{\frac{1}{4}}\int d LISP\, \mbarsquare
=\Eone\Etwo\,\sigma\vrel,
\label{w:eq}
\ee
where $d\,LISP$ in this case takes the form
\be
d\,LISP=(2\pi)^2\delta^4(\pone+\ptwo-\kone-\ktwo)
{\frac{d^3\kone}{(2\pi)^3 2 \Eone}} {\frac{d^3\ktwo}{(2\pi)^3 2
\Etwo}},
\ee
where $E_{1,2}=\vec{k}_{1,2}^2 + \mF^2$,
and
$\mbarsquare$ is the square of the reduced
matrix element for the annihilation process
${\bar\chi}\chi\ra{\bar F} F$
summed over the spins of the final-state particles and averaged over
the spins
of the initial particles~\cite{swo}.

Denote
\be
f\equiv\mbarsquare.
\label{fdef:eq}
\ee
In general $f=f(\vecpone,\vecptwo,\veckone,\vecktwo)$. The integral
Eq.~(\ref{w:eq}) can be conveniently evaluated in the centre-of-mass
frame (CM)
in which $\vecptwo=-\vecpone$ and $\vecktwo=-\veckone$.
After a few elementary steps one obtains
\be
w(s)= {\frac{1}{2^6\pi}} \sqrt{1-\frac{4\mF^2}{s}}
\int_{-1}^{1} d\!\cth f(\vecktwo=-\veckone,
|\veckone|=\sqrt{s/4-\mF^2}).
\label{wcm:eq}
\ee

Using the kinematic relation between the Mandelstam variables
\be
t=(\pone-\kone)^2=\mchi^2+\mF^2
-{\frac{s}{2}}\left[1-\sqrt{1-\frac{4\mchi^2}{s}}
\sqrt{1-\frac{4\mF^2}{s}}\cth\right],
\ee
one can express Eq.~(\ref{wcm:eq}) as
\be
w(s)= {\frac{1}{2^5\pi}} \frac{1}{s \sqrt{1-4\mchi^2/s}}
\int_{\tminus(s)}^{\tplus(s)} dt\, f(s,t),
\label{wst:eq}
\ee
where
$\tpm\equiv t(\cth=\pm1)=\tzero\pm\deltat$,
\be
\tzero(s)= \mchi^2 + \mF^2 - \frac{s}{2}
\label{tzero:eq}
\ee
and
\be
\deltat(s)= \frac{s}{2}
\sqrt{1-\frac{4\mchi^2}{s}} \sqrt{1-\frac{4\mF^2}{s}}
\label{deltat:eq}
\ee
Notice that one can always express $f$ as a function of the
Mandelstam
variables $s$ and $t$ only~\cite{swo}. In particular, $u$ can be
eliminated
by using the relation $s+t+u=2\mchi^2+2\mF^2$. Furthermore, in
calculating the
relic density it is convenient to introduce~\cite{swo}
\be
z\equiv {\frac{s}{4\mchi^2}}
\label{zdef:eq}
\ee
in terms of which Eq.~(\ref{wst:eq}) can be simply rewritten as
\be
w(z)= {\frac{1}{2^7\pi\mchi^2}} \frac{1}{z \sqrt{1-z}}
\int_{\tminus(z)}^{\tplus(z)} dt \,f(z,t).
\label{wzt:eq}
\ee

In order to calculate the relic abundance of $\chi$'s one needs to
solve the
Boltzmann (rate) equation.
The actual quantity that appears in the rate equation is the
thermally averaged product $\langle \sigmav \rangle$, which is
usually
approximated by $a+bx$, as mentioned above. One of the main results
of
Srednicki, \etal~\cite{swo}, was to show that
\bea
a&=&{\frac{1}{\mchi^2}} w(z=1)\\
b&=&{\frac{1}{\mchi^2}} \left[{\frac{3}{2}}{\frac{dw(z=1)}{d\!z}}
-3w(z=1)\right].
\label{abw:eq}
\eea

While these formulae look deceptively simple, the actual calculations
can be, and in practice {\em are}, very cumbersome and often
virtually
unmanageable. For example, in a relatively simple case of calculating
the
interference term between
two $t$-channel amplitudes (with the masses of the exchanged
particles
denoted by $\mu_1$ and $\mu_2$) one needs to evaluate several
integrals of the
type
$
\int_{\tminus}^{\tplus} dt\,t^n (t-\mu_1^2)^{-1}(t-\mu_2^2)^{-1}
$,
and in general in calculating $w(z)$ one has to compute a whole
multitude of
more complicated integrals. Once such a (lengthy) expression for
$w(z)$ is
found, one needs to next take a derivative $dw/d\!z$ which in general
leads to
even more cumbersome formulae. Additional highly non-trivial
computational
complications arise when the masses of the final state particles are
not equal.

Below I show that a great deal of these
difficulties can be avoided. In fact, one can avoid performing any
integrals {\em completely}.
First, one can always conveniently express~\cite{swo} $a$ and $b$ in
terms of
``reduced" variables $\azero$ and $\bzero$
\bea
a&=&\sqrt{1-\mF^2/\mchi^2}\, \azero, \label{azerodef:eq}\\
b&=&\sqrt{1-\mF^2/\mchi^2}\left\{
\left[-3+\frac{3}{4}\frac{\mF^2/\mchi^2}{(1-\mF^2/\mchi^2)}\right]
\azero
+  \bzero\right\}.
\label{bzerodef:eq}
\eea
I will show below that $\azero$ and $\bzero$ can be written in a
simple and
elegant form as
\bea
\label{azerow:eq}
\azero&=&{\frac{1}{2^5\pi\mchi^2}} f(z=1),  \\
\label{bzerow:eq}
\bzero&=&{\frac{1}{2^5\pi\mchi^2}}\left[
-3\mchi^2\frac{\partial f(z=1)}{\partial t}
+\mchi^2(\mchi^2-\mF^2)\frac{\partial^2 f(z=1)}{\partial t^2}
+\frac{3}{2}\frac{\partial f(z=1)}{\partial z}
\right]
\eea
and $f(z=1)$ should be understood as $f(z=1, \tzero(z=1))$,
\etc\ Equations~(\ref{azerow:eq}) and~(\ref{bzerow:eq}) are the main
result of
this Letter. They can also be readily generalized to the case of
unequal
final-state masses~\cite{lrinprep}. Higher order terms of the
expansion can
also be easily derived.
It is clear that the whole
procedure of calculating $\azero$ and $\bzero$ has now been reduced
to
merely writing down $\mbarsquare$,
substituting all the variables in $\mbarsquare$ in terms of
$z$ and $t$ and next taking a few relatively simple derivatives. In
fact, one
can easily do all these steps entirely with the help of any
advanced algebraic program. The truly difficult part of computing
$\azero$ and
$\bzero$ - performing complicated integrations in deriving $w(z)$ -
has been
completely eliminated.

In order to prove Eqs.~(\ref{azerow:eq}) and~(\ref{bzerow:eq}) notice
that for any regular function $g(z,t(z))$ and its integrand
$G(z)=\int dt\, g(z,t(z))$ one can show that
\bea
\lim_{z\ra 1} \left[ \frac{1}{\sqrt{1-z}}
G(z)\vert_{\tminus(z)}^{\tplus(z)}\right]&=&
\lim_{z\ra 1} \left[\frac{1}{\sqrt{1-z}}\int_{\tminus(z)}^{\tplus(z)}
dt\,
g(z,t(z))\right] \nonumber \\
&=& 4\mchi^2\sqrt{1- \mF^2/\mchi^2 }\, g(z=1,\tzero(z=1))
\label{gidentity:eq}
\eea
where I have used
$\lim_{z\ra 1} \tpm=\tzero(z=1)$, $\lim_{z\ra 1}
\deltat=2\mchi^2\sqrt{1-\mF^2/\mchi^2}\sqrt{1-z}$
and
\bea
\label{gexpand:eq}
\lim_{z\ra 1} G(z,\tpm) = \lim_{z\ra 1}
\Bigl[
&&G(z,\tzero)\pm g(z,\tzero)\deltat +
\frac{1}{2} \frac{\partial g(z,\tzero)}{\partial t} (\deltat)^2
\\
&&\pm \frac{1}{6} \frac{\partial^2 g(z,\tzero)}{\partial t^2}
(\deltat)^3 +
{\cal O}((\deltat)^4)
\Big]. \nonumber
\eea
(In deriving Eq.~(\ref{bzerow:eq}) one needs to
apply the formula~(\ref{gidentity:eq}) twice (for $f$ and $\partial
f/\partial
z$), and
keep terms up to $(\deltat)^3$ only in  the expansion of $f$ in order
to cancel
the term
$(z-1)^{-3/2}$ which appears in $dw/dz$ in Eq.~(\ref{wzt:eq}).)
Applying these formulae to Eqs.~(\ref{wzt:eq}),~(\ref{azerodef:eq}),
and~(\ref{bzerodef:eq}) one obtains
Eqs.~(\ref{azerow:eq}) and (\ref{bzerow:eq}) after a few simple
steps.

As an example, consider the process $\chi\chi\ra ZZ$ in the Minimal
Supersymmetric Standard Model (MSSM) with $\chi$ being the lightest
neutralino.
The two $s$-channel diagrams
include the exchange of the scalars $\hl$ and $\hh$, while the four
neutralinos $\chi^0_i$ ($i=1,4$) (including $\chi\equiv\chi^0_1$) are
exchanged in the $t$- and $u$-channels. Since neutralinos are
Majorana
particles, Eq.~(\ref{fdef:eq}) now reads
\bea
f&=&{\frac{1}{4}}\sum_{helicity}^{}\sum_{spin}^{} |{\cal
M}(\chi\chi\ra ZZ)|^2
={\frac{1}{4}}\sum_{helicity}^{}\sum_{spin}^{} |{\cal M}_{s}-{\cal
M}_{t}+{\cal
M}_{u}|^2 \\ \nonumber
&=& f^{(ss)} - f^{(st)} + f^{(su)} + f^{(tt)} - f^{(tu)} + f^{(uu)}.
\label{fzz:eq}
\eea

After expressing the external momenta in terms of $t$, $u$, and $z$
one finds, \eg,
\bea
f^{(tt)}(z,t)&=& {\frac{1}{8}}\sum_{helicity}^{}\sum_{spin}^{} |{\cal
M}_{t}|^2
\nonumber \\
&=&
{\frac{g^4}{\costhwfourth}}
\sum_{i,j=1}^{4}
( O_{1i}^{\prime\prime L})^2 ( O_{1j}^{\prime\prime L})^2
{\frac{1}{(t-\mi^2)(t-\mj^2)}} \sum_{k=0}^{4} \sum_{l=0}^{1}
f^{(tt)}_{kl} t^k
z^l,
\label{fttpartial:eq} \\
f^{(tu)}(z,t)&=& 2\Re\,
({\frac{1}{8}}\sum_{helicity}^{}\sum_{spin}^{} |{\cal
M}_{t}^{\dagger}{\cal M}_{u}|)   \nonumber \\
&=&
{\frac{g^4}{\costhwfourth}}
\sum_{i,j=1}^{4}
( O_{1i}^{\prime\prime L})^2 ( O_{1j}^{\prime\prime L})^2
{\frac{1}{(t-\mi^2)(u-\mj^2)}} \sum_{k=0}^{4} \sum_{l=0}^{2}
f^{(tt)}_{kl} t^k
z^l.
\label{ftupartial:eq}
\eea
It is also clear that $f^{(uu)}(z,u)=f^{(tt)}(z,t)$.
In the convention used here $O_{ij}^{\prime\prime L}= -\frac{1}{2}
(N_{i3}N_{j3} - N_{i4}N_{j4} )$ (in the basis
$(\bino,\wino,\hinob,\hinot)$),
the matrix $N_{ij}$ is real and the neutralino masses can be either
positive or
negative.
The expressions for the coefficients $f^{(tt)}_{kl}$ and
$f^{(tu)}_{kl}$,
introduced above, are rather lengthy and will be given
elsewhere~\cite{lrinprep}.
I will neglect here also the contribution from the $s$-channel
exchange~\cite{lrinprep}.

Finally, one obtains
\be
\azero= {\frac{g^4 (\mx^2 - \mz^2)}{4\pi\costhwfourth}}
\sum_{i,j=1}^{4}
{\frac{ ( O_{1i}^{\prime\prime L})^2 ( O_{1j}^{\prime\prime L})^2
}{\deli\delj}}
\label{azerofinal:eq}
\ee
and
\bea
\bzero&=& {\frac{g^4}{64\pi\costhwfourth \mz^4}}
\sum_{i,j=1}^{4}
{\frac{ ( O_{1i}^{\prime\prime L})^2 ( O_{1j}^{\prime\prime L})^2
}{\deli^3\delj^3}} \times 	\nonumber \\
&&
\Bigl[
\deli^2\delj^2 \left(
48\mi\mj\mx^4 + 32\mi\mx^5 +
    32\mj\mx^5 + 32\mx^6 \right.
    \nonumber \\
&&-
    48\mi\mj\mx^2\mz^2 -
    16\mi\mx^3\mz^2 -
    64\mj\mx^3\mz^2 +
    32\mx^4\mz^2 \nonumber \\
&&+
    36\mi\mj\mz^4 +
    11\mi\mx\mz^4 +
    29\mj\mx\mz^4 +
    12\mx^2\mz^4  +
    32\mz^6
\left.
    \right)
\nonumber \\
&&
+2\deli^2\delj \left(
    4\mi\mj\mx^6
-  24\mi\mj\mx^4\mz^2
+  10\mi\mx^5\mz^2 \right.
\nonumber \\
&&-  22\mj\mx^5\mz^2
+  20\mx^6\mz^2
+  20\mi\mj\mx^2\mz^4
+  2\mi\mx^3\mz^4
\nonumber \\
&&+  46\mj\mx^3\mz^4
+  5\mx^4\mz^4
- 12\mi\mx\mz^6
- 24\mj\mx\mz^6
- 25\mx^2\mz^6
\left.
\right)
\nonumber \\
&&
+2\deli\delj^2 \left(
  28\mi\mj\mx^6
- 24\mi\mj\mx^4\mz^2
- 58\mi\mx^5\mz^2
\right.
\nonumber \\
&&-  26\mj\mx^5\mz^2
+ 44\mx^6\mz^2
- 4\mi\mj\mx^2\mz^4 +
    82\mi\mx^3\mz^4
\nonumber \\
&&+
    38\mj\mx^3\mz^4
    - 13\mx^4\mz^4
-  24\mi\mx\mz^6
-  12\mj\mx\mz^6 - 31\mx^2\mz^6
\left.
\right)
\nonumber \\
&&
+16\deli\delj\left(
4\mi\mj\mx^8  -
    4\mi\mj\mx^6\mz^2 -
    6\mi\mx^7\mz^2
\right.
    \nonumber \\
&&-
    6\mj\mx^7\mz^2 +
    4\mx^8\mz^2 -
    4\mi\mj\mx^4\mz^4 +
    12\mi\mx^5\mz^4
\nonumber \\
&&+
    12\mj\mx^5\mz^4
-   5\mx^6\mz^4
+
    4\mi\mj\mx^2\mz^6 -
    6\mi\mx^3\mz^6
\nonumber \\
&&-
    6\mj\mx^3\mz^6
-   2\mx^4\mz^6
+ 3\mx^2\mz^8
\left.
\right)
\nonumber \\
&&
+32(\deli^2+\delj^2)\left(
\mx^6\mz^4 - 2\mx^4\mz^6 + \mx^2\mz^8
    \right)
\Bigr]
\label{bzerofinal:eq}
\eea
where $\Delta_{i,j}\equiv \mz^2-\mchi^2-m_{\chi^0_{i,j}}^2$.

These expressions reduce nicely to the Eqs.~(3.19.a) and (3.19.b) of
Ref.~\cite{os} in the limit in which $\chi$ becomes an almost pure
anti-symmetric higgsino $\hinoa\equiv 1/\sqrt{2}\, (0,0,-1,1)$
(corresponding in the MSSM to $\mtwo\gg\mu,\mz$~\cite{os}).
In this limit the second lightest neutralino $\neuttwo$ is an almost
pure
symmetric higgsino $\hinos\equiv 1/\sqrt{2}\, (0,0,1,1)$ and it is
almost
mass-degenerate with $\chi$,
$|m_{\chi,\neuttwo}|\simeq \mu (1\mp \epsilon)$ and $\mneuttwo\simeq
-\mchi$.
In Ref.~\cite{os} the contribution from the two heavier neutralinos,
which
in this limit are almost pure gauginos, has been neglected.
However, as has been pointed out by Drees and Nojiri~\cite{dn}, they
have in this limit non-negligible higgsino components which also
contribute to
the considered process. The expressions~(\ref{azerofinal:eq}) and
(\ref{bzerofinal:eq}) are free from this problem because they are
valid
for any type of neutralino and include contributions from all four
exchanged neutralinos.

The method presented here has also been tested on a number of other
cases (\eg,
$\chi\chi\ra (Z,\hl,\hh,\ha) \ra {\bar f} f$), for which simple
analytic
expressions are available. A complete set of relevant
expressions for the MSSM case will be presented
elsewhere~\cite{lrinprep}.

\bigskip

\noindent
{\em Note Added:}  After this work had been completed and submitted
to
a journal, I was made aware of
Ref.~\cite{jim} in which the expansion coefficients
for $\sigmav$ were
derived using a different method, without specifying
a thermal averaging procedure. When these expressions are applied to
the method
of thermal averaging of Ref.~\cite{kt}, which
does not include
relativistic corrections~\cite{swo,gg}, they become
different from the expression presented here by a factor
$b-b_x=-\frac{3}{2}a$~\cite{msprivate}.
\section*{Acknowledgments}
I would like to thank the organizers of the Workshop on Electroweak
Interactions at the Institute for Theoretical Physics at Santa
Barbara, where part of this project has been done, for their kind
hospitality.
I am also indebted to M.~Srednicki for discussions on the subject and
to M.~Drees, M.~Nojiri, and J.~Wells for pointing out a few
corrections.



\end{document}